\documentclass[twocolumn,amsmath,amssymb,floatfix,superscriptaddress,,prb]{revtex4-2}
\usepackage{bm,graphicx,url,epsf,color}
\usepackage{amssymb}
\usepackage{amsthm}
\usepackage[colorlinks=true,linkcolor=blue,citecolor=blue]{hyperref}
\usepackage{comment}

\begin{document}

\title{Anyonic exchange in a beam splitter}

\author{Christophe Mora}
\affiliation{Universit\'e Paris Cit\'e, CNRS,  Laboratoire  Mat\'eriaux  et  Ph\'enom\`enes  Quantiques, 75013  Paris,  France}

\begin{abstract}
The exotic braiding of anyons is certainly the most tantalizing aspect of fractional quantum Hall states. Although braiding is usually thought as a two-dimensional adiabatic manipulation, the braiding phase can also be captured in one dimension in an out-of-equilibrium setting. We discuss here to what extend a beam splitter
reveals the braiding phase when excited with voltage or current pulses. We identify two main physical mechanisms that govern the sign and the size of the output cross-correlations: the non-linearity, characterized by the tunneling exponent, and the presence of holes in the incoming beams, related to the braiding phase. We show how the incoming signals form an environment for the beam splitter (akin to the dynamical Coulomb blockade effect)
and thus interpret the mixing not as a collision of particles, but as a collision or interference of waves. We illustrate the physical picture with various examples of excitations for integer and fractional Hall states and show the emergence of genuine antibunching statistics when mixing dense voltage pulses.
\end{abstract}

\maketitle

{\bf Introduction} Anyons exhibit spectacular exchange and braiding statistics that distinguish them unambiguously 
from the conventional fermions and bosons~\cite{Halperin1984,arovas1984,nayak2008}. The predictions for the properties of anyons in the various fractional quantum Hall states are still awaiting an exhaustive experimental confirmation~\cite{heiblum2020}. Fractional charges have been revealed in several experiments including shot noise measurements~\cite{Saminadayar1997,DePicciotto1997}, finite-frequency emission noise~\cite{bisognin2019}, photo-assisted noise~\cite{kapfer2019} (see also Ref.~\cite{willett2009}). Chargeless quantized thermal transport has also been observed~\cite{Banerjee2017,Banerjee2018,Srivastav2019,Srivastav2021} but it is only recently~\cite{Nakamura2020} that a measure of the anyonic braiding phase has been realized in a Fabry-Perot geometry at filling $1/3$.

In addition and following theoretical proposals, a recent series of experiments~\cite{Bartolomei2020,glidic2022,glidic2022cross,ruelle2022,lee2022} has explored the mixing of fractional chiral edge states through a beam splitter, or quantum point contact (QPC), and measured the output cross-correlations. The prediction~\cite{rosenow2016}, confirmed quantitatively~\cite{Bartolomei2020} so far at filling $1/3$, is that the cross-correlations depend on the anyonic braiding phase and thus serve as a signature for non-trivial exchange statistics. Andreev scattering~\cite{kane2003} have also been demonstrated at strong transmission~\cite{glidic2022} or two-particle interferometry~\cite{Taktak2022}. Other theoretical proposals have followed with the mixing of non-abelian anyons~\cite{lee2022-2}, unconventional cross-correlations in the integer case with emulated fractional charges~\cite{morel2021,idrisov2022} or Hong-Ou-Mandel interferences to detect anyonic statistics~\cite{ronetti2018,jonckheere2022} or interactions~\cite{rebora2020,acciai2022}.

In this work, we investigate the processes of tunneling at a beam splitter and unravel the ingredients behind negative cross-correlations. Originally, they were intuitively understood as a deviation for anyons from fermionic antibunching during collisions~\cite{rosenow2016}. This explanation is however not transparently related to the analytical prediction as collisions are infrequent in the regime of interest of dilute anyonic beams. A corpuscular description is less appropriate than a wave description for the mixing of beams~\cite{lee2022-2}.
We show that a straightforward description emerges in the energy representation where the incoming beams provide an environment for the beam splitter, analoguous to the standard description of dynamical Coulomb blockade~\cite{ingoldnazarov1992,Altimiras2007,Hofheinz2011}. The environment is responsible for quasiparticle, but also hole, tunneling which, combined with the non-linearity of transport, result in negative cross-correlation.
After proving the equivalence between an upstream QPC and a random voltage bias to form dilute anyons, we consider various forms of pulses for integer filling. We show that even pulses quantized to a single unit charge can produce holes and thus lead to negative cross-correlations. The exception is the Lorentzian pulse which eludes the formation of holes due to its peculiar analytical structure. Finally, we determine the output correlations in the regime of dense voltage pulses. We find positive cross-correlations as an indicator of fermionic antibunching, distributing evenly the incoming electrons. \\





\begin{figure*}
\centering\includegraphics[width=2.\columnwidth]{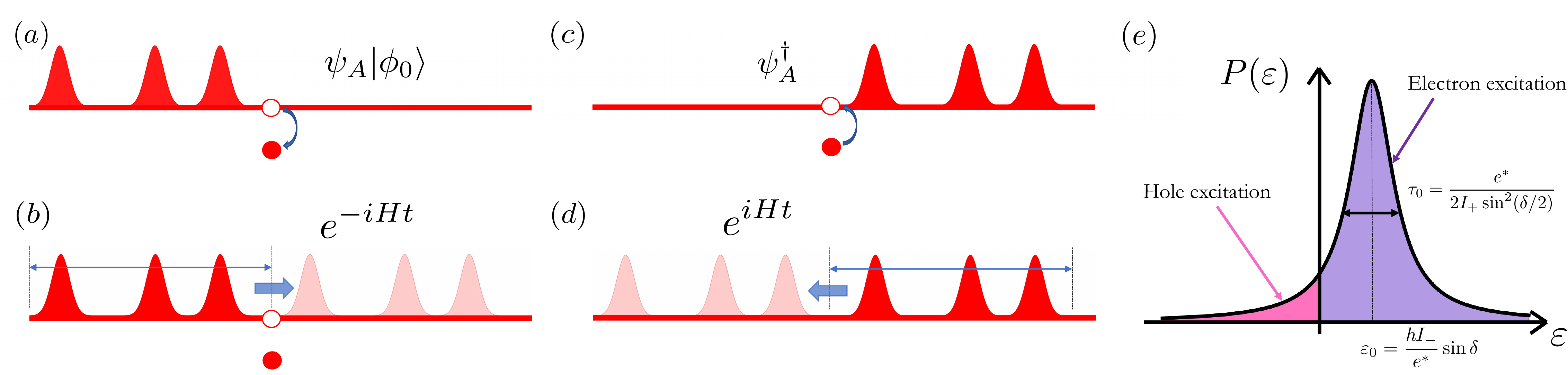}
\caption{(a-d) The quasiparticle correlator ${\cal C}_{\rm A} (t)$ is identified with the overlap~\eqref{overlap}. The state $| \phi_1 (t) \rangle$ is the result of four consecutive processes: (a) a quasihole is created at the position $x$ on the edge channel, (b) the states is evolved with the Hamiltonian $H$ during time $t$, translating $m$ anyons across the quasihole, (c) the quasihole at $x$ is (destroyed) filled, (d) a backward time evolution occurs, moving the $m$ anyons back across the former position of the quasihole. The total braiding phase $m \, \delta$ is acquired with respect to the unperturbed state $|\phi_0\rangle $ with the poissonian probability $p(m)$.  (e) Environment function at the beam splitter, see Eq.~\eqref{environmentfunction}. The right part ($\varepsilon>0$) induces quasiparticle tunneling from $A$ to $B$, the left part ($\varepsilon<0$) quasiparticle tunneling from $B$ to $A$ or, equivalently quasiholes from $A$ to $B$.  \label{fig1}}
\end{figure*}

{\bf Anyonic pulses on a chiral edge} Before investigating the signal mixing at a quantum point contact, we discuss the injection of quasiparticles on a quantum Hall edge with carrier velocity $v$.
For concreteness, we focus on the fillings $\nu=1$ for the integer case, and $\nu=1/3$ for the (abelian) fractional state. 
At equilibrium and zero temperature, the edge is characterized by the time-dependent correlator~\cite{schiller2022}
\begin{equation}\label{correlator1}
    \langle \psi_A^\dagger(x,t)\psi_A(x,0)\rangle = \frac{1}{(0^+ +i t)^{\nu} } \equiv {\cal C}_{\rm eq} (t) 
\end{equation}
for the quasiparticle creation operator $\psi_A^\dagger$ at any position $x$ along the edge. The random injection of anyons can be realized with an upstream QPC, as proposed in Refs.~\cite{kane2003,rosenow2016,lee2019,lee2020} and experimentally realized in Refs.~\cite{Bartolomei2020,glidic2022,glidic2022cross,ruelle2022,lee2022}. The source QPC couples weakly to a voltage-polarized auxiliary channel and injects dilute anyons with a poissonian distribution. In the presence of this excitation, the quasiparticle correlator $ {\cal C}_{\rm A} (t)  \equiv  \langle \psi_A^\dagger(x,t)\psi_A(x,0)\rangle $ of Eq.~\eqref{correlator1} changes into 
\begin{equation}\label{correlator2}
 {\cal C}_{\rm A} (t)  =  {\cal C}_{\rm eq} (t) \times \begin{cases}  e^{-  \frac{I_A |t|}{e^*} \left( 1 - e^{i \delta \, {\rm sgn} (t)}  \right)} \quad |t| \gg \tau \\[2mm]  1 + \frac{2 i \pi I_A t}{e} \qquad \qquad \quad |t| \ll \tau \end{cases}
\end{equation}
$\tau = \hbar/(e V)$ is the temporal width of each anyon set by the voltage drop $V$ across the upstream QPC.
$I_A$ is the average electrical current carried by the anyons, $e^* = \nu e$ the anyon charge and $\delta = 2 \pi \nu$ the phase accumulated after the braiding of two anyons. The long-time exponential decay of Eq.~\eqref{correlator2} is valid when $\delta$ is not a multiple of $2 \pi$. 

The injection of dilute anyons does not necessarily require an upstream QPC but can also be obtained with a purely classical, but random, time-dependent voltage, details are given in  appendix~\ref{append:A}. Applied with a metallic contact, the voltage $V_{\rm ex} (t) = \sum_j V_0 (t-t_j)$ is a sum of identical pulses, of temporal width $\tau$, normalized to $\int d t \, V_0(t) = h/e$ to excite a single anyon. The anyon injection times $t_j$ are randomly chosen with a poissonian distribution, meaning that injection events are fully independent. As shown in appendix~\ref{append:B}, the quasiparticle correlator obtained after averaging over the injection distribution precisely recovers Eq.~\eqref{correlator2} in both limits. 
The specific shape of the elementary pulse $V_0(t)$ does not matter for the asymptotic behaviours. The phase $\delta$ is moreover tunable by varying the pulse integrated voltage $\int d t \, V_0(t)$, see Eq.~\eqref{charge-individual}, without changing the scaling exponent $\nu$ in Eq.~\eqref{correlator2}.

The physics of anyon braiding is already encompassed in the quasiparticle correlator of Eq.~\eqref{correlator2}. In order to see this more clearly, we can rewrite the long-time behaviour of Eq.~\eqref{correlator2} as~\cite{morel2021,lee2022,lee2022-2}
\begin{equation}\label{sumpoi}
  {\cal C}_{\rm A} (t)  =  \sum_{m\ge 0} p(m) \,  {\cal C}_{\rm eq} (t)e^{i   m \delta}
\end{equation}
where $p(m) = (\bar{m}^m / m!)e^{- \bar{m}}$, $\bar{m} = I_A t / e^*$ is the poissonian probability to have $m$ anyons (or voltage pulses) within the time interval $t$. The correlator ${\cal C}_{\rm A} (t)$ also takes the form of an overlap between two states 
\begin{equation}\label{overlap}
{\cal C}_{\rm A} (t) = \langle \phi_0 | \phi_1 (t) \rangle
\end{equation}
$|\phi_0 \rangle$ is the ground state at zero temperature simply dressed by the excitation pulses.  $|\phi_1 (t)\rangle = e^{i H t} \psi_A^\dagger e^{-i H t} \psi_A |\phi_0 \rangle$ follows from a forward and a backward Hamiltonian evolution where a quasihole is respectively present or absent as pictured in Fig.~\ref{fig1}(a-d).
During the forward time evolution, $m$ anyons, with probability $p(m)$, pass the position $x$ of the quasihole. The same $m$ anyons revert back to their initial positions in the backward evolution but in the absence of a quasihole.
Apart from the equilibrium contribution  ${\cal C}_{\rm eq}$ the overlap thus measures the phase difference between the two situations with and without the local (anyon) quasiparticle as if the propagating anyons were encircling the local anyon. Each propagating anyon braid with the local one and the braiding phases add up to $m \, \delta$ as expressed in Eq.~\eqref{sumpoi}. Despite the fact that anyons move chirally in one dimension of space, time evolution and the presence/absence of a local hole reinstate a braiding process. \\





{\bf Energy representation and environment} It is instructive to switch to an energy representation where a more intuitive picture of tunneling will emerge. We split the correlator of Eq.~\eqref{correlator2} as ${\cal C}_{\rm A} (t) = {\cal C}_{\rm eq} (t) P_A(t)$ and Fourier transform all terms,
\begin{equation}\label{fouriertransform}
    {\cal C}_{\rm A} (\varepsilon) = \int_{-\infty}^{+\infty} d \varepsilon' \, {\cal C}_{\rm eq} (\varepsilon-\varepsilon') P_A(\varepsilon') =  {\cal C}_{\rm eq} \ast P_A (\varepsilon),
\end{equation}
where $\ast$ stands for the convolution product
and $P_A(\varepsilon)$ incorporates the anyon braiding.
In the integer case $\nu=1$, ${\cal C}_{\rm eq}(\varepsilon)$ is simply the Fermi distribution or $\theta (-\varepsilon)$ at zero temperature. 
For $\nu=1/3$ (see appendix~\ref{append:C}), ${\cal C}_{\rm eq}(\varepsilon) = (|\varepsilon|^{\nu-1}/\Gamma(\nu)) \theta (-\varepsilon)$ and is non-vanishing also solely for positive energies. Hence ${\cal C}_{\rm eq}(\varepsilon)$ plays  the role of the quasiparticle distribution in the ground state. The probability distribution $P_A (\varepsilon)$ is normalized by 
$\int d \varepsilon P_A(\varepsilon) = P_A(t=0)=1$.
In analogy with the dynamical Coulomb blockade effect, $P_A(\varepsilon)$ describes the exchange of energy between the chiral edge and its environment, here the bias voltage excitation. $P_A(\varepsilon)$ represents the probability to emit (absorb) an energy $\varepsilon$ to the environment for $\varepsilon<0$ ($\varepsilon>0$). Hence, the energy emission (absorption) creates holes (quasiparticles) in Eq.~\eqref{fouriertransform} in comparison with the equilibrium distribution ${\cal C}_{\rm eq} (\varepsilon)$ of the ground state. \\

{\bf Tunneling at the QPC}  
\begin{figure}
\centering\includegraphics[width=.9\columnwidth]{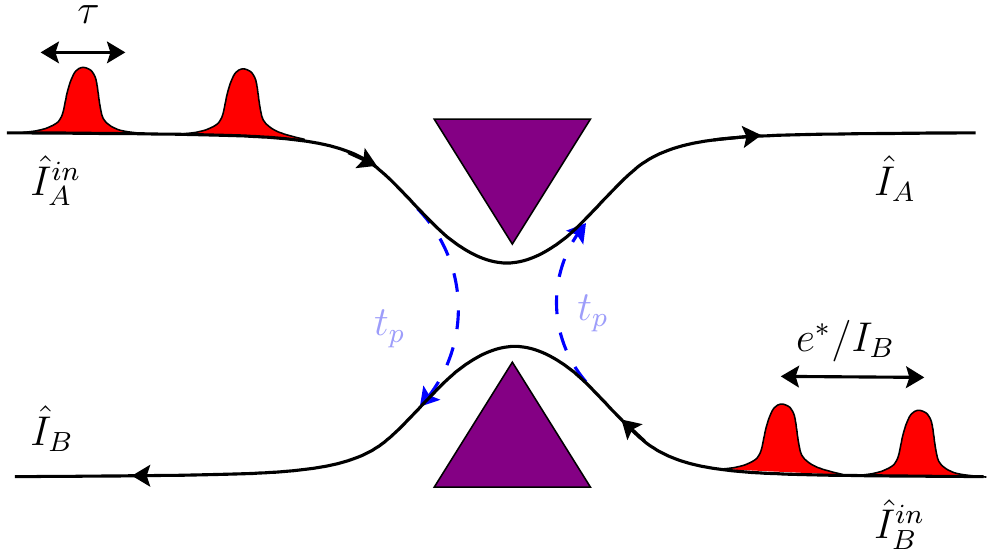}
\caption{Setup of the beam splitter. Two (fractional) chiral edge states, denoted $A$ and $B$, are mixed with a quantum point contact of transmission amplitude $t_P$. The input signals are injected either with weakly reflecting upstream QPCs or by utilizing random voltage sources as discussed in the text. The pulses have a width $\tau$ and a mean temporal distance $e^*/I_A$ (or $e^*/I_B$) between them. They are independent and their arrival statistics is poissonian. \label{fig2}}
\end{figure}
We now examine the mixing of two dilute anyonic beams, characterized by the environment (injection) functions $P_A (t)$ and $P_B (t)$, at a QPC. The setup is illustrated in Fig.~\ref{fig2}. At weak coupling, the tunneling of quasiparticles is modeled by the Hamiltonian $H_T = t_P  \tau_c^{\nu-1} \psi_B^\dagger \psi_A + {\rm h.c.}$ where the fields are taken at the position of the QPC ($\tau_c$ is a short-time cutoff and $t_P$ is dimensionless). Two parallel poissonian processes can occur: a quasiparticle may tunnel from $A$ to $B$ or vice-versa. The corresponding tunneling rates $w_{\rm AB}$ and $w_{BA}$ are calculated using the Kubo formula and the quasiparticle correlator Eq.~\eqref{correlator2} as recalled in appendix~\ref{append:C}. In the energy representation, the result takes however an intuitive form
\begin{equation}\label{tunnelingAB}
    w_{\rm AB/BA} = w_0 \int \int d\varepsilon d \varepsilon' \,  {\cal  C}_{\rm eq} (\varepsilon-\varepsilon')  {\cal \bar C}_{\rm eq} (\varepsilon)  P (\pm \varepsilon')
\end{equation}
where we introduce $w_0 = \frac{T_P}{2 \pi} \tau_c^{2 \nu-2}$, $T_P = 4 \pi^2 |t_P|^2$.
$P(\varepsilon)$ is the energy Fourier transform of $P(t) = P_A(t) P_B^*(t)$. ${\cal C}_{\rm eq} (\varepsilon)$ and  ${\cal \bar C}_{\rm eq} (\varepsilon)  \equiv {\cal C}_{\rm eq}(-\varepsilon)$ act as quasiparticle and quasihole reservoirs. The tunneling rate $w_{\rm AB}$ in Eq.~\eqref{tunnelingAB} involves a quasiparticle in $A$, a quasihole in $B$ and an energy transfer from (or to) the environment for $\varepsilon'$ positive (negative). At zero temperature,  ${\cal C}_{\rm eq} (\varepsilon) \sim \theta(-\varepsilon)$ and energy conservation implies that tunneling events from $A$ to $B$ are caused only by the $\varepsilon>0$ side of $P$ while only the complementary $\varepsilon<0$ side triggers tunneling from $B$ to $A$. This is illustrated in Fig.~\ref{fig1}(e).

The Fano factor of the quasiparticle tunneling through the QPC is given by 
\begin{equation}\label{fano1}
    F = \frac{w_{\rm AB}+w_{\rm BA}}{|w_{\rm AB}-w_{\rm BA}|}
\end{equation}
as shown in appendix~\ref{append:C}.
Particularly, for a function $P(\varepsilon)$ with a support only for $\varepsilon>0$ or $\varepsilon<0$, one of the two rates $ w_{\rm AB/BA}$ identically vanishes and the Fano factor $F=1$ is obtained, corresponding to poissonian shot noise. This is for instance the case with a DC bias polarization $V$ across the QPC with, in that case, $P(\varepsilon) = \delta(\varepsilon + e^* V)$. $V>0$ yields tunneling events from $A$ to $B$ and no tunneling from $B$ to $A$ while $V<0$ gives the opposite. This is true both in the integer and fractional cases.

A transparent physical picture can be given in the integer case $\nu=1$ where
the tunneling rates and Fano factor can readily be related to the excess numbers of electrons and holes~\cite{dubois2013-2,rech2017}. We consider that electrons on the $A$ channel see both environment functions $P_{A/B}$ at the QPC so that Eq.~\eqref{fouriertransform} becomes $ {\cal C}_{\rm A} = {\cal C}_{\rm eq} \ast P$. The excess number of electrons is thus $N_e = \int d \varepsilon \, {\cal C}_{\rm A} (\varepsilon)   \theta (-\varepsilon)$ and for holes $N_h = \int d \varepsilon [1- {\cal C}_{\rm A} (\varepsilon)]  \theta (\varepsilon)$. Using these definitions, we find $w_{\rm AB} = w_0 N_e$, $w_{\rm BA} = w_0 N_h$ and the Fano factor
\begin{equation}\label{fano2}
    F = \frac{N_e+N_h}{|N_e-N_h|}.
\end{equation}
The presence of both excess electrons and holes in the distribution leads to a Fano factor larger than one as both tunnel processes are activated.
Although we have shown the connection rigorously for $\nu=1$ only, the effect is still qualitatively valid in the fractional case $\nu=1/3$. In particular, when ayonic pulses are injected only via $A$, that is for $P_B=1$, the excess anyons cause tunneling from $A$ to $B$ but the presence of holes also yields tunnel events from $B$ to $A$ to fill these holes, {\it against the excitation current}.


Coming back to the correlator Eq.~\eqref{correlator2}, we obtain a Lorentzian form for the environment function, represented in Fig.~\ref{fig1}(e),
\begin{equation}\label{pefunc}
    P (\varepsilon) = \frac{\hbar}{\pi \tau_0} \frac{1}{(\varepsilon- \varepsilon_0)^2 + \hbar^2/\tau_0^2}
\end{equation}
with $\varepsilon_0 = \hbar \frac{I_-}{e^*} \sin \delta$ and $1/\tau_0 = \frac{I_+}{e^*} [1-\cos \delta]$, having defined $I_{\pm} = I_A\pm I_B$ in terms of the incoming current fluxes $I_A$ and $I_B$ (both positive). Interestingly, $\varepsilon_0/e^*$ acts as an effective bias voltage whereas $\tau_0$ sets a lifetime that broadens the emission (absorption) energy distribution $P(\varepsilon)$. Assuming $\varepsilon_0$ and $\tau_0$ to be independent - in practice they are not -  one finds $P(\varepsilon) = \delta(\varepsilon-\varepsilon_0)$ as $\tau_0 \to +\infty$, precisely as a DC bias voltage directly applied across the QPC. With the expression~\eqref{pefunc} having support for all energies, both quasiparticles and holes are injected in the two channels and the Fano factor in Eq.~\eqref{fano1} is necessarily larger than $1$. Inserting the expression of the environment function~\eqref{pefunc} into Eq.~\eqref{tunnelingAB} and performing the integral, we arrive at
\begin{equation}\label{transferrate}
    w_{\rm AB/BA} = {\cal N}_0 {\rm Re} \, \left[ e^{-i \pi \nu} \left(I_+ \mp  \frac{i I_-}{\tan \delta/2} \right) ^{2 \nu -1} \right],
\end{equation}
in agreement with Ref.~\cite{rosenow2016,lee2022}. In the particular case where $I_B=0$, {\it i.e.} there is only current injection from one side, the Fano factor reduces to $F= \cot (\pi \nu) \cot [ (\pi-\delta)(1/2-\nu) ]$. Numerically, the value for the fractional case $\nu = 1/3$, and $\delta =2 \pi/3$, is $F \simeq  3.274$, as recently predicted and experimentally verified~\cite{lee2022}. \\

{\bf Noise cross-correlations} The tunneling processes at the QPC and the way the two incoming signals couple is well characterized by measuring cross-correlations in the output beams. In a purely (classical) corpuscular description, the incoming poissonian quasiparticles arrive at the QPC and are randomly distributed among the two output lines. Since the resulting output beams are still poissonian, the sum of noise auto-correlations is preserved by the QPC and charge conservation thus implies vanishing cross-correlations. In other words, the two output signals are poissonian and uncorrelated.

Non-zero cross-correlations hence measure deviations from the classical poissonian reference. Of course, collisions of quasiparticles in a corpuscular description provide such a deviation by including the quantum statistics of the collision, bunching or anti-bunching, but collisions are rare for dilute beams and can be made arbitrarily small by reducing the incoming currents. Therefore, the deviations that matter in the dilute limit arise from correlated events of quasiparticle or hole tunnelings at the QPC, as we will now determine. 
A small fluctuation $\delta \hat I_A^{\rm in}$ in the input current compared to its average value $I_A$ affects the tunnel current from $A$ to $B$,
\begin{equation}\label{deltaIT}
    \delta \hat I_T = e^* \left( - \frac{\partial w_{AB}}{\partial I_A} \delta \hat I_A^{\rm in} + \frac{\partial w_{BA}}{\partial I_A} \delta \hat I_A^{\rm in} \right)
\end{equation}
and correlate the incoming and outgoing currents. A similar expression is derived for the current fluctuation $\delta \hat I_B^{\rm in}$ in channel $B$. Using these relations and the tunneling rates, the normalized cross-correlations in the symmetric case $I_A = I_B$ take the form
\begin{equation}\label{presult}
{\cal P} \equiv \frac{S_{\rm AB}}{e^* I_+ \partial_{I_-} \langle \hat I_T \rangle} = 1 - \frac{1}{\lambda_1-\lambda_2}   
\end{equation}
where we have introduced the two logarithmic derivatives
\begin{equation}\label{logderivative}
    \lambda_{1} = \frac{\partial \ln w_{\rm AB}}{\partial \ln I_{A}}, \qquad \quad  \lambda_{2} = \frac{\partial \ln w_{\rm AB}}{\partial \ln I_{B}}.
\end{equation}
Details are deferred to appendix~\ref{append:D}.
$\lambda_1$ characterizes the non-linearity of quasiparticle tunneling, it is one for a linear dependence on the incoming current, the exponent in the case of a power law. $\lambda_2$ accounts for the reverse current where tunneling occurs against the excitation signal. 
For $\nu=1/3$ and $\delta = 2 \pi/3$, we obtain from Eq.~\eqref{transferrate}, $\lambda_{1/2} = (\nu-1/2) [1 \mp \tan(\pi \nu) \cot(\delta/2)]$, yielding $\lambda_1=0$, $\lambda_2 = -1/3$ and the normalized noise cross-correlations ${\cal P}=-2$, experimentally confirmed in Ref.~\cite{Bartolomei2020}.

$\lambda_1 <1$ and $\lambda_2>0$ both decrease ${\cal P}$ to negative values. Intuitively, cross-correlations result from two opposite effects~\cite{rosenow2016,morel2021}. The first one is the granular noise of  tunneling ($\delta \hat I_T$) which anti-correlates the output channels $A$ and $B$, the second one comes from the shot noise of the incoming signal: a current excess $\delta \hat I_A^{\rm in}$ splits into the $A$ and $B$ channels and contributes positively to cross-correlations. The former effect is strengthened by the reverse (hole) flow which increases the level of tunneling noise, the latter is weakened by less-than-linear transfer efficiency $\partial_{I_A} w_{\rm AB} < w_{\rm AB}/I_A$, or $\lambda_1<1$.

We see that collisions of corpuscular anyons do not emerge as an adapted description as the neither linear nor quadratic dependence on the incoming currents indicates. Instead, we have an interference of the incoming waves that produces tunneling or, loosely speaking, a collision of waves.


It is important to stress at this point that the Lorentzian form Eq.~\eqref{pefunc} corresponds to the long-time exponential decay of Eq.~\eqref{correlator2}. $P(t)$ has also a short to intermediate time dependence which depends on the details of the pulse, correcting Eq.~\eqref{pefunc} at high energies. For pulses quantized to $\delta = 2 \pi$, the exponential decay is absent in $P(t)$ and the tunneling at the QPC is governed by the full time dependence as discussed thereafter. For $\nu=1$ and $\delta \ne 2 \pi$, the average tunnel current $\langle \hat I_T \rangle$ is determined by the short-time behaviour and cross-correlations by long times, as shown in the Supplementary and Ref.~\cite{morel2021}. Long times dominate both observables when $\nu \ne 1$ and $\delta \ne 2 \pi$.\\


{\bf Voltage pulses in the integer case}  
To gain further insight into the operating philosophy of the QPC beam splitter, we explore the integer case $\nu=1$ in more detail. It has already been discussed that excitations of the chiral channels can lead to negative cross-correlations ${\cal P} <0$, either by injecting fractional charges with an upstream QPC combined with a metallic island~\cite{morel2021}, or from the mere Coulomb interaction between two channels at filling factor two~\cite{idrisov2022}. We note that, similarly to the fractional case discussed after Eq.~\eqref{correlator2}, the QPC and island combination of Ref.~\cite{morel2021} can be replaced by a classical time-dependent  voltage consisting of identical fractionally normalized but random pulses. They lead to arbitrarily large and negative values of the cross-correlations, logarithmically divergent with the beam diluteness. From our discussion above, they are easily understood in terms of holes created by the voltage pulses exciting each only a fraction of an electron. These holes amplify the reverse current and increase the level of noise produced by the QPC which anticorrelates the output channels $A$ and $B$.

We now show that, even for a train of quantized voltage pulses, holes are generically created, thereby inducing negative cross-correlations. Only Lorentzian pulses preclude hole creation~\cite{dubois2013-2} and yield vanishing cross-correlations. The applied voltage on each chiral edge has the form $V_{\rm ex} (t) = \sum_j V_0 (t-t_j)$, where the injection times $t_j$ are randomly chosen and independent. The quantization of each pulse is characterized by the phase $\phi_0 (t) \equiv \frac{e}{\hbar} \int^t_{-\infty} d t' \, V_0 (t')$ which winds from $0$ to $2\pi$ over the pulse duration. As shown in appendix~\ref{append:A}, the environment function is obtained as $P_A (t) = e^{-(I_A/e) g(t)}$ with 
\begin{equation}\label{functiong}
    g(t) = \int_{-\infty}^{+\infty} d t' \, \left( 1 - e^{i \left [ \phi_0 (t-t') - \phi_0 (-t') \right]}  \right),
\end{equation}
with the short-time behavior $g(t) \simeq 2 i \pi t$.
The function $g(t)$ remains bounded when the pulse is quantized, otherwise it increases linearly with $t$ as given in Eq.~\eqref{correlator2}. We prefer to discuss the tunneling rates in the time domain: $w_{\rm AB} = (w_0/2 \pi) \int d t \, G(t) [{\cal C}_{\rm eq} (t)]^2$, with $G(t) = (I_+/e) g_1(t) + i (I_-/e) g_2 (t)$. $g_{1/2}$ denote the real and imaginary parts of $g(t)$. The expression for $w_{\rm BA}$ is the same except that $G$ is replaced by its complex conjugate $G^*$. After a few algebraic manipulations, we obtain
\begin{equation}
    w_{\rm AB/BA} = \frac{T_P}{2 e} \big( \pm I_- + z_0 I_+ \big),
\end{equation}
where $z_0 = \int d t g_1 (t)/(2 \pi^2 t^2)$ is evaluated numerically. Using the logarithmic derivatives of Eqs.~\eqref{logderivative}, we find $\lambda_1 = (z_0+1)/2 z_0$ and $\lambda_2 = (z_0-1)/2 z_0$ and thus the noise cross-correlations 
\begin{equation}
{\cal P} = 1-z_0
\end{equation}
The Fano factor of Eq.~\eqref{fano1} is moreover given by $F = z_0 I_+/|I_-|$, {\it i.e.} $F_0 = z_0$ when a single channel is excited, {\it e.g.} $I_B=0$.
\begin{table}[]
\centering
\begin{tabular}{ |c|c|c| } 
\hline
 $e V_0 (t)/h$ & $\quad F \, \, (z_0) \quad$ & $\qquad {\cal P} \qquad$ \\[1mm] 
 \hline \hline 
$ f_1 (t) \equiv \dfrac{\tau}{\pi} \dfrac{1}{t^2+\tau^2}$ & $1$ & $0$ \\[5mm]  
$f_2 (t) \equiv \dfrac{1}{2 \tau} \dfrac{1}{\cosh^2 (t/\tau)}$ & 1.04541 & -0.04541 \\[5mm]
$f_3 (t) \equiv \dfrac{1}{2 \tau \Gamma(5/4)} e^{-t^4/\tau^4}$ & 1.09888 &  -0.09888 \\[5mm]
$ \dfrac{f_1(t)+f_1 (t- 20 \tau) + f_1 (t+20 \tau)}{3}$ & 1.49781 &  -0.49781 \\[3mm]
 \hline
\end{tabular}
\caption{Cross-correlations and Fano factors for different injection signals, all normalized to excite one electron per pulse, $\int d t V_0 (t) = h/e$. The last function is denoted $f_4$ and all the corresponding integrated phases are shown in Fig.~\ref{fig3}.}
\label{table1}
\end{table}

\begin{figure}
\centering\includegraphics[width=.9\columnwidth]{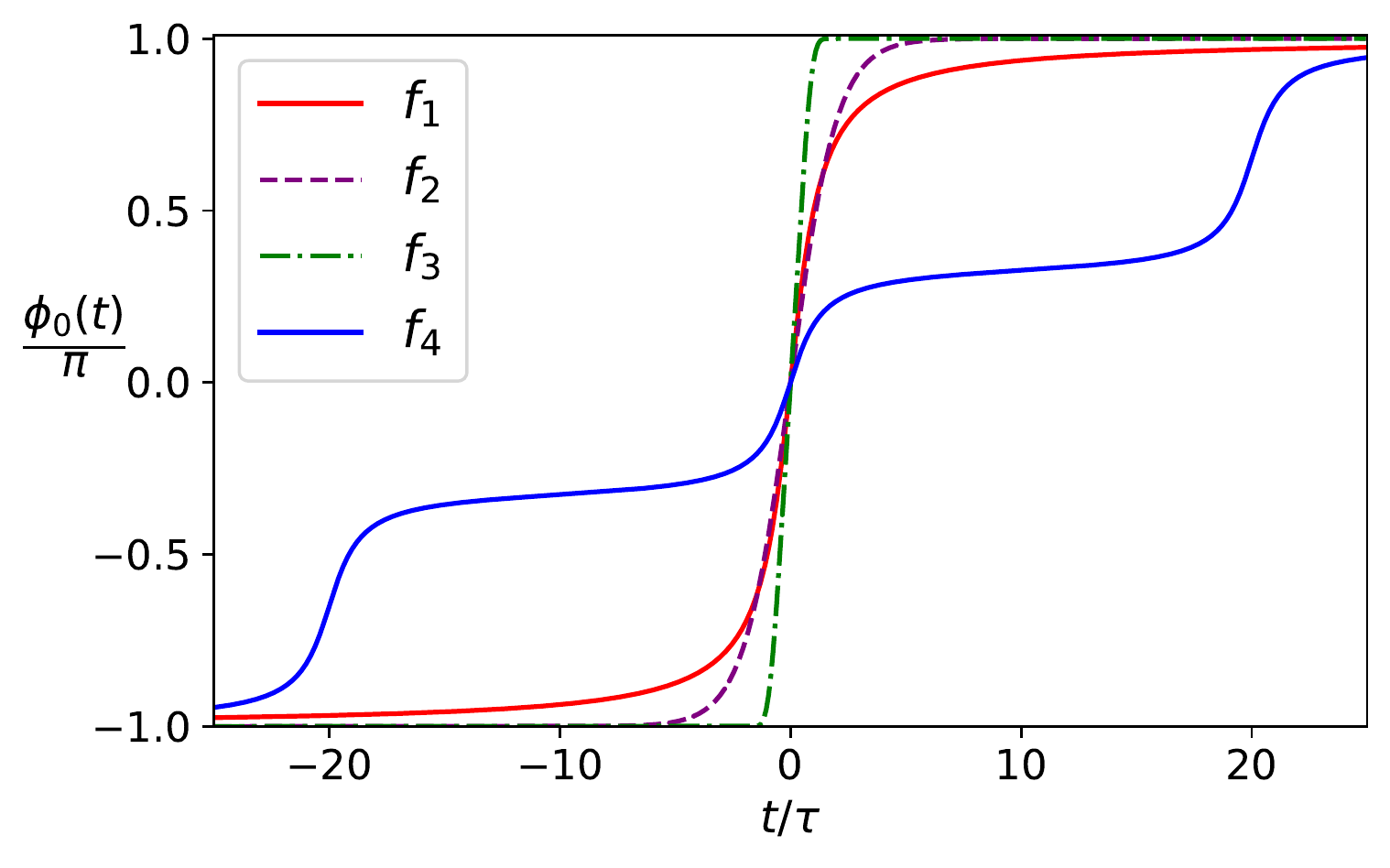}
\caption{Phase accumulated for an elementary pulse,  $\phi_0 (t) \equiv \frac{e}{\hbar} \int^t_{-\infty} d t' \, V_0 (t')$.  The different curves refer to the voltage excitations listed in table~\ref{table1}. \label{fig3}}
\end{figure}

The result $z_0=1$ is obtained for a Lorentzian pulse, yielding vanishing cross-correlations ${\cal P}=0$ and the Fano factor $F=1$. Mathematically, this is understood by noticing the absence of poles in the upper-half plane for the function $g(t)$, akin to an excitation of the Fermi sea with only electrons and no hole. As a consequence of this pole structure, the time integrals for $w_{\rm AB/BA}$ disentangle the two chiral channels: $w_{\rm AB} \sim I_A$  and $w_{\rm BA} \sim I_B$, yielding an absence of reverse (hole) current at the QPC, $\lambda_1 = 1$ and $\lambda_2=0$. For all other elementary quantized pulses, holes are created and $z_0 >1$, ${\cal P}<0$ and $F>1$. A few examples are compiled in table~\ref{table1} and the corresponding phases are plotted in Fig.~\ref{fig3}.

For all functions corresponding to a single step-increase of the phase $\phi_0 (t)$, {\it i.e.} a single pulse for $V_0 (t)$, the deviation to the Lorentzian form is not sufficient to induce sizeable cross-correlations. {\cal P} thus remains small in absolute value. However, a separation into multiple steps (or multiple pulses)~\cite{yue2021} gives significant cross-correlations. 
Even though the total signal remains quantized to a single electron excitation, cross-correlations can be made arbitrarily large (and negative) by increasing the number of pulses and the distance between them, making the link with the logarithmic divergence found for purely fractional pulses~\cite{morel2021}. \\

{\bf Antibunching for dense wavepackets} Still keeping a voltage excitation made of quantized integer (unit) pulses, we extend the analysis beyond the dilute regime. We consider a time separation between pulses $e/I_A$ that becomes comparable with their width $\tau$ and show that antibunching of fermions occurs as wavepackets start overlapping and collisions take place. We also consider random Lorentzian pulses as they allow for a full analytical treatment. By virtue of the independence of individual voltage pulses, the incoming beam is always poissonian, even in the dense regime, as discussed in appendix~\ref{append:A}. The full environment function is 
\begin{equation}\label{environmentfunction}
P (t) =  \exp \left( - \frac{I_A}{e}  g(t) - \frac{I_B}{e} g^*(t) \right)
\end{equation}
where, using the general expression Eq.~\eqref{functiong}, one finds
$g(t) =  4 \pi \tau t/(t+2 i \tau)$
for Lorentzian voltage pulses. When the pulses are dilute with the small parameter $I_{A/B} \tau/e \ll 1$, one can expand the full environment function $P(t) \simeq 1- (I_A/e) g(t) - (I_B/e) g^* (t)$. An aspect we already mentioned is that $g(t)$ is analytic in the upper-half complex plane and $g^*(t)$ in the lower-half. As a consequence, the time integral for the rate $w_{AB}$ has zero contribution from the latter and the result depends on $I_A$ only. The conclusion is reversed for $w_{BA}$ which depends linearly and only on $I_B$. In a sense, the two incoming signals do not mix and trigger each a different tunneling process for dilute beams. 

\begin{figure}
\centering\includegraphics[width=.9\columnwidth]{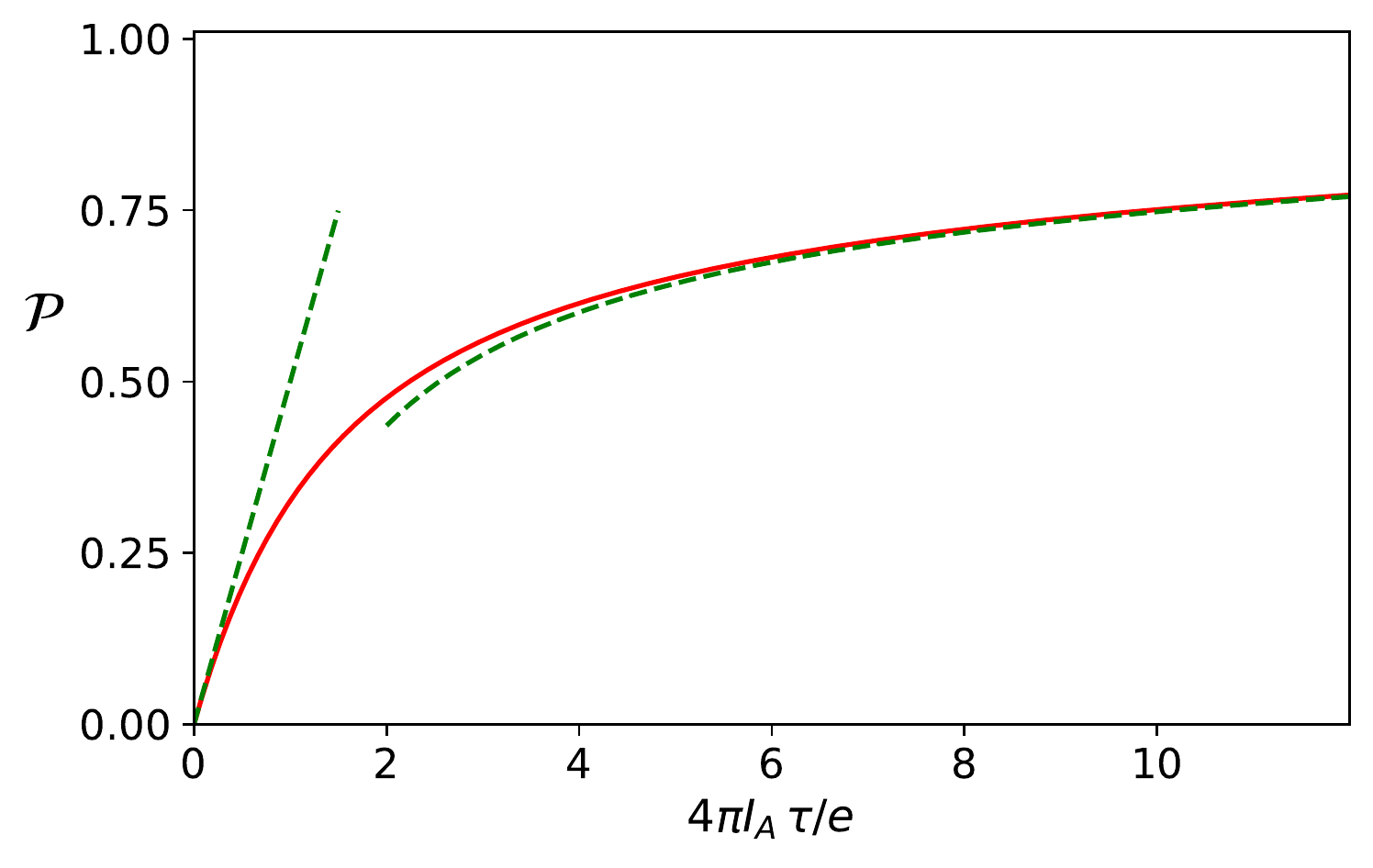}
\caption{Renormalized cross-correlation ${\cal P}$ (red line), defined in Eq.~\eqref{presult}, as a function of the dilution parameter $\alpha = 4 \pi I_A \tau /e$ for $I_A=I_B$, and compared to the asymptotical behaviors $\alpha/2$ and $1-\sqrt{2/(\pi \alpha)}$ (green dashed lines). $\alpha$ is the ratio between the duration of a single pulse and the mean distance between consecutive pulses. \label{fig4}}
\end{figure}

The absence of mixing for Lorentzian pulses is no longer true outside the dilute limit as $P(t)$ in Eq.~\eqref{environmentfunction} has poles both in the upper and lower half-planes. An analytical formula (see Supplementary) for the tunneling rate is obtained for balanced beams $I_A = I_B$,
\begin{equation}
    \frac{w_{AB}}{w_0} = \int  \frac{d t}{2 \pi} \, P(t) \,  [{\cal C}_{\rm eq} (t)]^2 = \pi \alpha e^{-\alpha} \left[ I_0 (\alpha) + I_1 (\alpha) \right]
\end{equation}
and $w_{BA} = w_{AB}$,
with the modified Bessel functions $I_0$ and $I_1$, and $\alpha \equiv 4 \pi I_A \, \tau/e$. Moreover, we also find $\langle \hat I_T \rangle = T_P I_-$ irrespective of the diluteness parameter $I_A \tau /e$. The final result for cross-correlations hence takes the form
\begin{equation}\label{crosscorrelation-dense}
    {\cal P} = 1 - e^{- \alpha} \left[ I_0 (\alpha) + I_1 (\alpha) \right],
\end{equation}
for balanced beams $I_A=I_B$, which interpolates between the dilute result ${\cal P} = 0$ for $\alpha \to 0$ and ${\cal P} = 1$ in the very dense limit, $\alpha \to +\infty$, see Fig.~\ref{fig4}.

Interestingly, the result of Eq.~\eqref{crosscorrelation-dense} shows that, in the regime where electron wavepackets, of temporal size $\tau$, begin to overlap, $\alpha \sim 1$, collisions take place between them at the QPC which tends to distribute them evenly between the output channels $A$ and $B$, thereby inducing positive cross-correlations. This democratic distribution is forced by the Pauli exclusion principle and corresponds to antibunching. Notably, in the setup considered here, the overlap of independent wavepackets is as probable for electrons in the same incoming channel as it is for electrons originating from opposite channels. \\

{\bf Conclusion and discussion} We have provided a comprehensive analysis of the beam splitter as an analyzer for the braiding phase in an abelian fractional quantum state. Focusing on the excitation with quantized and dilute current pulses carrying the elementary fractional charge, we have shown that the braiding phase already enters the out-of-equilibrium field correlator and can  later be extracted after passing through a beam splitter realized with a QPC. 
We have identified two main physical effects that govern the amplitude and sign of the noise cross-correlations: the non-linearity of quasiparticle tunneling at the QPC and the presence of holes in the energy distribution of the incoming signals. The physics is the most transparent in the energy domain where the incoming pulses provide an out-of-equilibrium environment triggering quasiparticle tunneling processes in both directions. Holes in particular favor charge transfers against the excitation current and drive cross-correlation towards negative values or the QPC Fano factor above $1$. We have confirmed this picture by exploring in detail integer quantum Hall states excited by voltage pulses, where negative cross-correlation directly relate to the hole production. Still in the integer case, we have also demonstrated that fermionic antibunching emerges as the distance between current pulses is reduced, resulting in positive cross-correlations. In this dense regime, collisions between fermionic wavepackets play a crucial role.

For fractional edge states, Lorentzian periodic pulses quantized to move the elementary  charge $e^*$ were shown in Ref.~\cite{rech2017} to produce particle-hole pairs, in contrast to levitons in the case~\cite{keeling2006}. With the expression of the environment function $P(\varepsilon)$ in Eq.~\eqref{pefunc}, we confirm quasiparticle-hole pair production for our random quantized Lorentzian pulses but also when an upstream QPC injects dilute quasiparticles as in Ref.~\cite{rosenow2016,Bartolomei2020,glidic2022,lee2022}, yielding a Fano factor larger than one and negative cross-correlations. Following Ref.~\cite{rech2017}, one could alternatively use Lorentzian pulses carrying the electron charge $e$ (and not $e^*$) to suppress particle-hole pairs in the incoming beams. In this interesting case, the environment function $P(t)$ does not decay at long time and can be expanded in powers of the incoming currents $I_A,I_B$ as in the integer case. As a result, the tunneling process from $A$ to $B$ is linear in the dilute regime, or $\lambda_1=1$, and the Lorentzian form of the individual pulses removes quasihole tunneling, $\lambda_2=0$. The cross-correlations are nevertheless not vanishing as in the integer case because the granularity of charge transport differs: $e$ in the incoming signal, $e^*$ for tunneling at the QPC. The result ${\cal P} = 1-(e^*/e)^2$ is eventually obtained from Eq.~\eqref{cross-correl}.

In summary, our work establishes a general framework for understanding the statistics of charge transfer at a beam splitter with arbitrary incoming signals in the integer or fractional quantum Hall regimes. With the use of the environment function, it can be generalized to other types of edge stage excitations, with AC voltages or additional dissipative effects. We also leave for future works a generalization to finite frequency measurements where the description in terms of an environment function could give a transparent picture.


\begin{acknowledgments}
We thank H.-S. Sim, F. Pierre, G. F\`eve, T. Morel, A. Anthore, J.-Y. Lee, D.C. Glattli and O. Maillet for fruitful discussions. This work was supported by the French National Research Agency (project SIMCIRCUIT, ANR-18-CE47-0014-01).
\end{acknowledgments}

\appendix
\section{Random series of voltage pulses}\label{append:A}
In the presence of a voltage excitation injected by a metallic contact, the quasiparticle field is dressed by a phase: $\psi_A \to \psi_A e^{- i \phi_A (t)}$, which integrates the applied potential. We consider for the excitation signal a sum of identical pulses $V_{\rm ex} (t) = \sum_j V_0 (t-t_j)$ where the time variables $t_j$ are chosen randomly and independently from each other. $V_0(t)$ gives the shape of a single pulse characterized by a temporal width $\sim \tau$. The total phase is given by a similar sum, $\phi_A (t) = \sum_j \phi_0 (t-t_j)$, with $\phi_0 (t) = (e^*/\hbar) \int^t_{-\infty} d t' \, V_0 (t')$.
We average over the random times $t_j$ the dressing of the field correlator $P_A (t) = \langle e^{i (\phi_A (t) - \phi_A(0))} \rangle$
which, using the independence of pulses, factorizes as 
\begin{equation}
P_A (t) = \prod_j P_0 (t) = [P_0 (t)]^{\cal N}.
\end{equation}
${\cal N} = T I_A/q$ is the number of pulses within a long temporal window $T$ used for the average $\langle \ldots \rangle_{t_j} = \frac{1}{T} \int_{-T/2}^{T/2} d t_j \ldots$, $I_A$ the average current and $q$ the charge carried by each individual pulse. $P_0 (t) = 1 - g(t)/T$ is the dressing of the correlator due to a single pulse, the expression of the function $g(t)$ has been given in Eq.~\eqref{functiong}. We have assumed here that $T \gg \tau$ such that each pulse is fully covered by the time integration in the average. We can finally take the limit of very large $T$,
\begin{equation}
    P_A(t) = \left( 1 - \frac{g(t)}{T} \right)^{(I_A/q) T} \xrightarrow[T \to +\infty]{} e^{- (I_A/q) \, g(t)},
\end{equation}
and the charge moved by an individual pulse is determined by the phase increment
\begin{equation}\label{charge-individual}
    \frac{q}{e} =  \frac{\phi_0 (+ \infty) - \phi_0 (-\infty)}{2 \pi} = \frac{e^*}{h} \int^{+\infty}_{-\infty} d t \, V_0 (t)
\end{equation}
Note that the randomness discussed here differs from the pseudorandom binary injection discussed in Ref.~\cite{glattli2018}.

\section{Fractional pulses and asymptotics}\label{append:B}

The short-time behaviour of $g(t)$ is obtained by expanding inside and down the exponential in Eq.~\eqref{functiong} with the result 
\begin{equation}
g(t) \simeq - i t \int_{-\infty}^{+\infty} d t' \partial_t \phi_0 (-t') = - i t \, \Delta \phi_0 
\end{equation}
Expanding also the environment function $P(t) = P_A (t) P_B^* (t)$ at short time and using Eq.~\eqref{charge-individual}, we obtain $P(t) \simeq 1 + 2 i \pi (I_- /e) t$ which is the desired result of Eq.~\eqref{correlator2}. The short-time behaviour is insensitive to the charge $q$ carried by each pulse but also remarkably to the anyon charge $e^*$.

The long-time behaviour depends sensitively on the pulse charge $q$. When $q$ is an integer, the time integral in Eq.~\eqref{functiong} concentrates around $t' \simeq 0$ and $t' \simeq t$ and remains finite as $t$ is large. Therefore $g(t) \sim {\cal O} (1)$, the precise value depending on details of the voltage pulse. On the contrary, for $q$ different from an integer, the time integral in Eq.~\eqref{functiong} has a support between $0$ and $t$ and the long-time behaviour emerges 
\begin{equation}\label{long-time}
g(t) \simeq |t| \left( 1 - e^{2 i \pi (q/e) {\rm sgn} (t)} \right)
\end{equation}
with the use of Eq.~\eqref{charge-individual}. For a quantized pulse with $\int d t \, V_0(t) = h/e$, the charge in Eq.~\eqref{charge-individual} is $q = e^* = \nu e$ and Eq.~\eqref{long-time} retrieves Eq.~\eqref{correlator2} from the main text. 

We note that the quantization of $q/e$ to an integer is quite generally related to a minimized production of holes in the energy distribution as discussed in Ref.~\cite{dubois2013,dubois2013-2}.

\section{Full counting statistics at the QPC}\label{append:C}

The statistics of charge transfer, or full counting statistics, in a QPC are known in the tunnel regime~\cite{levitov2004}. The generating function $\chi_t (\lambda) = \sum_q P_t (Q) e^{i \lambda Q/e^*}$ determines all cumulants for the probability distribution $P_t (q)$ to transfer a total charge $Q$ during a time interval $t$ (from $A$ to $B$). Writing the tunnel Hamiltonian as $H_T = \hat J_{AB} + \hat J_{BA}$ with $\hat J_{AB} = t_P  \tau_c^{\nu-1} \psi_B^\dagger \psi_A$ and  $\hat J_{BA} = \hat J_{AB}^\dagger$, the result at weak transmission is
\begin{equation}\label{FCS}
\ln \chi_t (\lambda) = t \left[ \left ( e^{i \lambda} - 1 \right) w_{AB} + \left ( e^{-i \lambda} - 1 \right)  w_{BA} \right],
\end{equation}
corresponding to two independent poissonian processes whose 
rates coincides with Kubo's formula
\begin{equation}
w_{AB} = \int_{-\infty}^{+\infty} d t \, \langle \hat J_{BA} (t) \hat J_{AB} (0) \rangle,
\end{equation}
and $A$ and $B$ are simply interchanged in $w_{BA}$. The computation of these rates thus involves products of field correlators Eq.~\eqref{correlator1}. In some cases, the energy representation is more convenient with, for instance,
\begin{equation}
    {\cal C}_{\rm eq}(\varepsilon) = \frac{1}{2 \pi \hbar} \int_{- \infty}^{+\infty} d t \,  {\cal C}_{\rm eq} (t) e^{-i \varepsilon t /\hbar}
\end{equation}
Using the equilibrium expression in Eq.~\eqref{correlator1}, the integral can be performed and ${\cal C}_{\rm eq}(\varepsilon)  = (|\varepsilon|^{\nu-1}/\hbar^\nu \Gamma(\nu)) \theta (-\varepsilon)$ as announced in the main text (we took $\hbar=1$ there for simplicity). $\Gamma (x)$ is the Euler Gamma function. With Eq.~\eqref{FCS}, the mean current and noise of the QPC are readily extracted, $\langle \hat I_T \rangle =  e^* (w_{AB} - w_{BA})$ and $S_T = \langle \delta \hat I_T \delta \hat I_T \rangle = (e^*)^2 (w_{AB} + w_{BA})$. The Fano factor is thus $F = S_T/(e^* \langle \hat I_T \rangle)$ yielding Eq.~\eqref{fano1} in the main text.

\section{Current conservation and cross-correlations}\label{append:D}

With the voltage excitation introduced in appendix~\ref{append:A}, the incoming beams have poissonian distributions and $\langle (\delta \hat I_A^{\rm in})^2  \rangle = q I_A$ where $q$ is the granular charge carried by each pulse. The same holds for the $B$ channel. Current conservation generally implies~\cite{Blanter2000,morel2021} 
\begin{equation}\label{current-conservation}
\langle (\delta \hat{I}^{\rm in}_A)^2  \rangle 
+ \langle (\delta \hat{I}^{\rm in}_B )^2  \rangle = 
\langle (\delta \hat{I}_A)^2 \rangle 
+ \langle (\delta \hat{I}_B)^2  \rangle +  2 S_{AB}
\end{equation}
with the cross-correlations $S_{AB} = \langle \delta \hat{I}_A \delta \hat{I}_B \rangle$, and $\hat I_{A/B} = \hat{I}^{\rm in}_{A/B} \mp \hat I_T$. If the output beams also have poissonian statistics with charge $q$, then current conservation automatically implies from Eq.~\eqref{current-conservation} an absence of cross-correlation $S_{AB} =0$.

In general, the cross-correlations are expressed as~\cite{rosenow2016,morel2021,lee2022-2}
\begin{equation}
    S_{AB} = - S_T + \langle \delta \hat I_T (\delta \hat I^{\rm in}_A - \delta \hat I^{\rm in}_B ) \rangle 
\end{equation}
where the tunnel current $\hat I_T$ and incoming currents are correlated, see Eq.~\eqref{deltaIT}. Using Eq.~\eqref{deltaIT}, its equivalent on the $B$ channel and the fact that incoming beams are poissonian, one obtains for the reduced cross-correlations $S_{AB}/q^2$,
\begin{equation}\label{cross-correl}
\begin{split}
    - \left( \frac{e^*}{q} \right)^2 & ( w_{AB} + w_{BA}) \\[2mm]
    + & \left( I_A \partial_{I_A} - I_B \partial_{I_B} \right) (w_{AB} - w_{BA}),
\end{split}
\end{equation}
eventually leading to Eq.~\eqref{presult} for balanced beams $I_A=I_B$ and $q=e^*$.

\bibliography{biblio.bib}

\onecolumngrid

\clearpage

\setcounter{figure}{0}
\setcounter{section}{0}
\setcounter{equation}{0}
\renewcommand{\theequation}{S\arabic{equation}}
\renewcommand{\thefigure}{S\arabic{figure}}

\onecolumngrid

\renewcommand{\thesection}{S-\Roman{section}}
\renewcommand{\theequation}{S-\arabic{equation}}
\renewcommand{\thefigure}{S-\arabic{figure}}


\section*{Supplementary material: Anyonic exchange in a beam splitter}



We detail here the computation of noise cross-correlations for dilute to dense interfering beams in integer quantum Hall edge states. The injection is done on both channels $A$ and $B$ with random series of Lorentzian pulses
\begin{equation}\label{signel-pulse}
    V_0 (t) = \frac{h}{e} \frac{\tau}{\pi} \frac{1}{t^2+ \tau^2},
\end{equation}
and the interference takes place at a quantum point contact (QPC). $\tau$ is the typical duration of one pulse.
For instance, the voltage applied upstream from the beam splitter (or QPC) in channel $A$ has the form
\begin{equation}\label{vex}
    V_{\rm ex} (t) = \sum_j \, V_0 (t-t_j) 
\end{equation}
where the injection times $t_j$ are chosen independently from each other. Each pulse moves a unit charge
\begin{equation}
q = \frac{e^2}{h} \int_{-\infty}^{+\infty} d t \, V_0 (t) = e,     
\end{equation}
{\it i.e.} a single electron. The specificity of the Lorentzian pulse, also called a Leviton, is that it excites solely an electron above the Fermi sea with no accompanying electron-hole pairs. The mean distance between pulses is $e/I_A$ where $I_A$ is the average current. The injected signal is a dilute train of well-separated pulses when $I_A \tau/e \ll 1$. On the contrary, the signal becomes dense when $I_A \tau/e \simeq 1$ (or $I_A \tau/e \simeq 1$) and individual pulses Eq.~\eqref{signel-pulse} start overlapping.

Exactly the same signal is injected in channel $B$. The integral of the voltage Eq.~\eqref{signel-pulse} gives the phase 
\begin{equation}\label{phasephi0}
\phi_0 (t) = \frac{e}{\hbar} \int_{-\infty}^t d t' \, V_0 (t') =  \pi + 2 \arctan (t/\tau)  
\end{equation}
and dresses the field correlator $\langle \psi_A^\dagger(x,t)\psi_A(x,0)\rangle$ as
\begin{equation}
e^{i [\phi_0 (t-t_j) - \phi_0(-t_j)]} = \frac{-t_j+i \tau}{-t_j-i \tau} \, \frac{t-t_j-i \tau}{t-t_j+i \tau},
\end{equation}
with a pole only in the lower half-plane for the variable $t$.
This last expression is averaged over the injection time $t_j$,
\begin{equation}\label{average}
\langle e^{i [\phi_0 (t-t_j) - \phi_0(-t_j)]} \rangle_{t_j} \equiv \frac{1}{T} \int_{-T/2}^{T/2} d t_j e^{i [\phi_0 (t-t_j) - \phi_0(-t_j)]}  \simeq  1 - \frac{g(t)}{T},
\end{equation}
where we take a very long time $T$ for the average, much larger than all relevant time scales, and
\begin{equation}
    g(t) = \int_{-\infty}^{+\infty} d t_j \left(1-  \frac{-t_j+i \tau}{-t_j-i \tau} \,  \frac{t-t_j-i \tau}{t-t_j+i \tau} \right) = \frac{4 \pi t \tau}{t+2 i \tau},
\end{equation}
which remains analytical in the upper half-plane, and with the property $g^*(t)=g(-t)$. We stress that the result  is bounded $|g(t)| \le 4 \pi \tau$.
The electron operator in channel $A$ is dressed by the phase
$\phi_A (t) = (e/\hbar) \int^t d t' \, V_{\rm ex} (t')$. With Eq.~\eqref{vex}, Eq.~\eqref{phasephi0} and Eq.~\eqref{average}, it multiplies the field correlator $\langle \psi_A^\dagger(x,t)\psi_A(x,0)\rangle$ by
\begin{equation}
    \prod_{j=1}^{I_A T/e} \langle e^{i [\phi_0 (t-t_j) - \phi_0(-t_j)]} \rangle_{t_j} = \left(1 - \frac{g(t)}{T} \right)^{I_A T/e} \simeq e^{-(I_A/e) \, g(t)}
\end{equation}
for $I_A T/e \gg 1$, $I_A T/e$ being the average number of individual pulses during the integration time $T$ needed for the average and eventually sent to infinity. The correlator for the electron field in channel $B$ is similarly dressed and
\begin{equation}
  \langle \psi_A^\dagger(x,t)\psi_A(x,0)\rangle \langle \psi_B (x,t) \psi_B^\dagger (x,0)\rangle = \left[ {\cal C}_{\rm eq} (t) \right]^2 \,  P(t) = \frac{e^{-(I_A/e) \, g(t) - (I_B/e) \, g^*(t)}}{(0^+ +i t)^{2}}.
\end{equation}
Using the Kubo formula, we arrive at the tunneling rates
\begin{equation}
   w_{AB} = w_0 \int_{-\infty}^{+\infty}  \frac{d t}{2 \pi} \, \frac{e^{-(I_A/e) \, g(t) - (I_B/e) \, g^*(t)}}{(0^+ +i t)^{2}}
\end{equation}
$I_A$ and $I_B$ are simply interchanged in the expression of $w_{BA}$. We first note that ($w_0 =  T_P/(2 \pi)$)
\begin{equation}\label{differencew}
    w_{AB} - w_{BA} = \frac{T_P}{4 \pi^2} \int_{\cal C}  d z \frac{P(z)}{z^2} = \frac{T_P}{4 \pi^2} (-2 i \pi) P'(t=0) = \frac{T_P I_-}{e} 
\end{equation}
where the contour ${\cal C}$ runs in the complex plane slightly above the real axis from $-\infty$ to $+\infty$ and then slightly below the real axis from $+\infty$ to $-\infty$. ${\cal C}$ is thus a closed contour that encircles a single pole at $z=0$. We have used the short-time expansion $P(t) \simeq 1 +  2 i \pi (I_- /e) t$ derived in the manuscript (appendix $B$). The result Eq.~\eqref{differencew} holds  regardless of the exciting potential and is not limited to the Lorentzian form of Eq.~\eqref{signel-pulse}. It does not even require that each individual pulse moves a quantized charge. However, it is only valid when the bare constituent are electrons, {\it i.e.} in the integer quantum Hall case, implying that the equilibrium propagator ${\cal C}_{\rm eq} (t)$ has a pole. In the fractional quantum Hall case, ${\cal C}_{\rm eq} (t)$ exhibits a branch cut at $t=0$ and Cauchy's theorem cannot be used to compute $w_{AB} - w_{BA}$. \\

We now wish to determine $w_{AB}$ in the symmetric case where $I_A = I_B$. In this regime, $w_{AB} = w_{BA}$. We need to evaluate the following integral
\begin{equation}
  w_{AB} = \frac{w_0}{2 \pi \tau} \, I(\alpha) \qquad \qquad   I (\alpha) = \int_{-\infty}^{+\infty}  \frac{d x}{(0^+ +i x)^{2}} \, e^{- \alpha \left(\frac{x}{2 i +x} + \frac{x}{-2 i +x} \right)},
\end{equation}
where $\alpha = 4 \pi I_A \tau/e$ is the diluteness parameter. A series expansion of the exponential leads to 
\begin{equation}
    I (\alpha) = \sum_{n_1=0}^{+\infty} \sum_{n_1=0}^{+\infty} \frac{(-\alpha)^{n_1+n_2}}{n_1 ! n_2 !} \, J(n_1,n_2),  
\end{equation}
with
\begin{equation}
    J(n_1,n_2) = \int_{-\infty}^{+\infty}  \frac{d x}{(0^+ +i x)^{2}} \left( \frac{x}{2 i +x} \right)^{n_1} \left( \frac{x}{-2 i +x} \right)^{n_2} = - 2 \pi \left( \frac{1}{2} \right)^{n_1+n_2} \binom{n_1+n_2-2}{n_2-1},
\end{equation}
and  $J(1,0) = -\pi$, $J(n_1\ge 2,0) = 0$, $J(0,n_2\ge 0) = 0$. Using the identities
\begin{equation}
    \sum_{n_1=1}^{n-1} \frac{1}{n_1 ! (n-n_1)! (n_1-1)! (n-n_1-1)!} = \frac{4^{n-1}}{\sqrt{\pi}} \frac{\Gamma(n-1/2)}{n ! (n-1)! (n-2)!}
\end{equation}
and 
\begin{equation}
\sum_{n=1}^{+\infty}  \frac{(-2 \alpha)^n \, \Gamma(n-1/2)}{n! (n-1)!} = -2 \sqrt{\pi} \alpha e^{-\alpha} \left[ I_0 (\alpha) + I_1 (\alpha) \right],
\end{equation}
we finally arrive at the desired result
\begin{equation}
    I (\alpha) = \pi \alpha e^{-\alpha} \left[ I_0 (\alpha) + I_1 (\alpha) \right] \qquad \qquad w_{AB} = w_{BA} = \frac{T_P I_A}{e} e^{-\alpha} \left[ I_0 (\alpha) + I_1 (\alpha) \right]
\end{equation}
announced in the main text. The cross-correlation are then given by
\begin{equation}
S_{AB} = -e^2 (w_{AB} + w_{BA}) + e I_+ \partial_{I_-} \langle \hat I_T \rangle.
\end{equation}
With $\langle \hat I_T \rangle = T_P I_-$, we obtain
\begin{equation}
    {\cal P} \equiv \frac{S_{\rm AB}}{e I_+ \partial_{I_-} \langle \hat I_T \rangle} = 1 - e^{- \alpha} \left[ I_0 (\alpha) + I_1 (\alpha) \right],
\end{equation}
also written as Eq.~(19) in the main text.

\end{document}